\documentclass[apjl]{emulateapj}

\usepackage{graphicx}

\def\es{}

\def\esss{}
\def\eess{  }
\def\eees{}
\def\mtt{  }
\def\corr{}

\def\mtt{ }

\def\mtta{}
\def\ggg{$\gamma$-ray }

\shortauthors{}

\begin{document}

\submitted{Published by Astrophysical Journal, Volume 765, 52 (Feb. 2013)}
\title{Variable gamma-ray emission from the Crab Nebula: Short Flares and Long ``Waves"}




\author{E.~Striani\altaffilmark{1,2,3}, M.~Tavani\altaffilmark{1,2,3},
V.~Vittorini\altaffilmark{1},
I.~Donnarumma\altaffilmark{1},
A.~Giuliani\altaffilmark{5},
G.~Pucella\altaffilmark{4},
A.~Argan\altaffilmark{1},
A.~Bulgarelli\altaffilmark{6},
S.~Colafrancesco\altaffilmark{11,12},
M.~Cardillo\altaffilmark{1,2},
E.~Costa\altaffilmark{1},
E.~Del Monte\altaffilmark{1},
A.~Ferrari\altaffilmark{8},
S.~Mereghetti\altaffilmark{5},
L.~Pacciani\altaffilmark{1},
A.~Pellizzoni\altaffilmark{9},
G.~Piano\altaffilmark{1},
C.~Pittori\altaffilmark{10},
M.~Rapisarda\altaffilmark{4},
S.~Sabatini\altaffilmark{1},
P.~Soffitta\altaffilmark{1},
M.~Trifoglio\altaffilmark{6},
A.~Trois\altaffilmark{9},
S.~Vercellone\altaffilmark{7},
F.~Verrecchia\altaffilmark{10}}


\altaffiltext{1} {INAF/IASF-Roma, I-00133 Roma, Italy}
\altaffiltext{2} {Dip. di Fisica, Univ. Tor Vergata, I-00133 Roma,
Italy}
\altaffiltext{3} {INFN Roma Tor Vergata, I-00133 Roma, Italy}
\altaffiltext{4} {ENEA Frascati,  I-00044 Frascati (Roma), Italy}
\altaffiltext{5} {INAF/IASF-Milano, I-20133 Milano, Italy}
\altaffiltext{6} {INAF/IASF-Bologna, I-40129 Bologna, Italy}
 \altaffiltext{7} {INAF-IASF Palermo, Palermo, Italy}
\altaffiltext{8} {CIFS-Torino, I-10133 Torino, Italy}
\altaffiltext{9} {INAF-Osservatorio Astronomico di
Cagliari, localita' Poggio dei Pini, strada 54, I-09012 Capoterra,
Italy}
\altaffiltext{10} {ASI Science Data Center, I-00044
Frascati(Roma), Italy}
\altaffiltext{11} {INAF - Osservatorio Astronomico di
Roma via Frascati 33, I-00040
Monteporzio, Italy.}
\altaffiltext{12} {University of the Witwatersrand,
Private Bag 3, 2054 South Africa}

\begin{abstract}

{\mtt Gamma-ray emission from the Crab Nebula has been recently
shown to be unsteady.}
 In this paper, we study the {\mtta flux and spectral} variability of the Crab
above 100 MeV on different timescales ranging from days to weeks.
{\mtt In addition to} the four main {\mtt intense and day-long}
flares detected by AGILE and Fermi-LAT between Sept. 2007 and
Sept. 2012, we {\mtt find evidence {\corr for week-long} and less
intense episodes of} enhanced gamma-ray emission that we call
``waves". {\mtt Statistically significant  ``waves'' show
timescales of 1-2 weeks, and can occur by themselves or in
association with shorter flares.} We present a refined {\mtt flux
and spectral} analysis of the Sept. - Oct. 2007 gamma-ray {\mtt
enhancement episode detected} by AGILE that shows {\mtt both
``wave'' and flaring behavior.}  {\mtt We extend our analysis to
the publicly available Fermi-LAT dataset and show that several
additional ``wave'' episodes can be identified. }
{\mtt We discuss the spectral properties of the September 2007
``wave''/flare event and show that the physical properties of the
``waves'' are intermediate between steady and flaring states.
Plasma instabilities inducing ``waves'' appear to involve spatial
distances $ l \sim 10^{16} \,$cm and enhanced magnetic fields
{\esss $B \sim (0.5 - 1)\,$}mG. Day-long flares are characterized
by smaller distances and larger local magnetic fields. {\mtta
{\corr Typically, the deduced total energy associated with the
``wave'' phenomenon  ($E_w  \sim 10^{42} \, \rm erg$, where $E_w$
is the kinetic energy of the emitting particles) is comparable with}
that associated to the flares, and can reach a few percent of the
total available pulsar spindown energy.  {\corr Most likely,
flares and waves are the product of the same class of plasma
instabilities that we show acting on different timescales and
radiation intensities.} }}

\end{abstract}



\maketitle


\section{Introduction}

The Crab Nebula (the remnant of a Supernova explosion witnessed by
Chinese astronomers in 1054)  {\mtt is powered by} a very powerful
pulsar (of period $P=0.33$ ms, and  spindown luminosity $L_{sd}
\simeq 5 \times 10^{38} \, \rm erg \, s^{-1} $) (see e.g., Hester
2008). {\mtt The pulsar} is energizing the whole system through
the interaction of the particle and wave output within the
surrounding Nebula {\corr(of average magnetic field $\sim 200 \mu$G)}.
The {\mtt resulting unpulsed emission} from
radio to gamma rays up to $100$ MeV is interpreted as synchrotron
radiation from {\mtt at least} two populations of
electrons/positrons energized by the pulsar wind and by
surrounding shocks or plasma instabilities (e.g.,
Atoyan \& Aharonian 1996, Meyer et al. 2010). The optical and
X-ray brightness enhancements observed in the inner Nebula, known
as ``wisps'', ``knots'', and the ``anvil'' aligned with the pulsar
``jet'' (Scargle 1969; Hester 1995, 2008; Weisskopf 2000),
show {\mtt flux} variations on timescales of weeks or months.
{\mtt On the other hand,} the {\mtt average} unpulsed emission
from the Crab Nebula was always considered essentially stable.
{\mtt  The surprising discovery by the AGILE satellite of variable
gamma-ray emission from the Crab Nebula in Sept. 2010
\cite{tavani1, tavani3}, and the Fermi-LAT confirmation
\citep{buehler, abdo2}  started a new era of investigation of the
Crab system. As of Sept. 2012 we know of four major gamma-ray
flares from the Crab Nebula detected} by
{\corr the AGILE Gamma-Ray Imaging Detector (GRID)} and Fermi-LAT:
\textit{(1)} the Sept-Oct. 2007 event,
\textit{(2)} the Feb. 2009 event,
\textit{(3)} the Sept. 2010, and \textit{(4)}  the
``super-flare'' event of Apr. 2011
(Buehler et al. 2011; Tavani et al.
2011b; Hays et al. 2011; Striani et al. 2011a; Striani et
al. 2011b; Buehler et al. 2012).

  {\mtta In this paper we address the issue of the gamma-ray
  variability of the Crab Nebula on different timescales ranging
  from days to weeks. We then enlarge the parameter space sampled
  by previous investigations especially for the search of
  statistically significant enhanced emission on timescales of 1-2
  weeks. Sect. 2 presents a brief overview of the current
  knowledge on Crab's main gamma-ray flares. Sect. 3 presents the results of
  a search of \ggg enhanced emission on timescales of weeks in the AGILE database.
  We also discuss in
  detail the Sept.-Oct. 2007 event detected by AGILE which
  shows a strong evidence of short timescale flaring as well as
  substantial emission on longer timescales of order of 1 week.
  Sect. 4 presents the results of our search for long and short
  timescale enhanced emission in the available Fermi-LAT data. In
  both
  the $\gamma$-ray telescopes data we find strong evidence of
  week-long \ggg enhanced emission episodes at intermediate peak intensities that we
  call ``waves''. Sect. 5 presents the physical implications of our
  findings in terms of a synchrotron emission model. We discuss in Sect. 6
  the main implications of our work.}

\begin{table*}
  \begin{center}
    \caption{Table of the \emph{flares} (F $\geq 700 \times 10^{-8} \rm \, ph \, cm^{-2} \, s^{-1}$)
  of the Crab Nebula found in the AGILE and Fermi data from Sept. 2007.}\label{tab1}
  \begin{tabular}{|c|c|c|c|c|c|c|c|c|}
    \hline
    & Name &  MJD & \textbf{$\tau_{1}$} (hr) & \textbf{$\tau_{2}$} (hr) & Peak Flux & $B (mG)$ & $\gamma^{\ast}$ ($10^{9}$) &  $l$ ($10^{15}$ cm)\\
    \hline
  2007  & $F_{1}$   & 54381.5   & {\eess $22\pm 11$}    & {\eess $10\pm 5$}  & $1000\pm 150 $     & $ 1.0 \-- 2.0 $     & $2.6\--4.8$    &$1.2\--3.6$\\
   (AGILE)  & $F_{2}$   & 54382.5   & {\eess $14\pm 7$}     & {\eess $6\pm 3$}   & $1400 \pm 200 $    &  {$\esss 1.1\-- 2.1$}    & ${\esss 2.3\--4.3}$   & $0.8\--2.2$\\
            & $F_{3}$   &54383.7    & {\eess $11\pm 5$}    & {\eess $14\pm7$}    & $900 \pm 150$            & $1.0 \-- 2.0$    & $2.6\--4.8$  &$0.8\--1.7$\\ \hline
  2009      & $F_{4}$   &54865.8    & {\eess $10\pm 5$}      & {\eess $20\pm 10$}   &  $700 \pm 140 $   & {$\esss 0.7\--1.3$}    & $2.6\--4.8$    &$0.6\--1.6$ \\
  (FERMI)   & $F_{5}$   &54869.2    & {\eess $10\pm 5$}   & {\eess $22\pm 11$}    & $830 \pm 90 $  & {$\esss 0.8\-- 1.4$} & $2.6\--4.8$      &$0.6\--1.6$\\
   \hline
   2010     & $F_{6}$   &55457.8    & {\eess $8\pm 4$}    & {\eess $22\pm 11$}      & $850 \pm 130 $    & $0.7\-- 1.3$    & $2.5\--4.7$    &$0.5\--1.3$\\
   (AGILE \& & $F_{7}$   &55459.8 & {\eess $6\pm 3$}    & {\eess $6\pm 3$}     & $1000 \pm 100 $   & $ 1.4\--2.6 $   & $2.6\--4.8$     &$0.3\--0.9$\\
      FERMI)      & $F_{8}$   &55461.9  & {\eess $19\pm 10$}  & {\eess $8\pm 4$}   & $750 \pm 110 $   &  {$\esss 0.8\--1.4$}    & $2.5\--4.8$   & $0.9\--3.1$ \\ \hline
  2011      & $F_{9}$   &55665.0    & {\eess $9\pm 5$}     & {\eess $9\pm 5$}        & $1480 \pm 80 $    & $ 1.2\--2.2 $   & $2.8\--5.0$    & $0.5\--1.5$ \\
  (FERMI \&  & $F_{10}$  &55667.3  & {\eess $10\pm 5$}  & {\eess $24\pm 12$}  & $2200 \pm 85 $  & {$\esss 1.3\--2.3$}  & ${\esss 2.7\--4.9}$  & $0.6\--1.6$\\
  AGILE) & & & & & & & &\\ \hline
    \hline
  \end{tabular}
    \end{center}
\noindent {\mtt The timescales $\tau_1$ and $\tau_2$ are the rise
and decay timescales of the flares modelled with {\eess an
exponential fit}
respectively. The {\esss
characteristic length} of the emitting region is {\eees deduced}
{\corr from the relation $l=c \, \delta \, \tau_1$}. The Lorentz factor
$\gamma^{\ast}$ characterizes the adopted model of the  accelerated
particle distribution function $d \, n/d \, \gamma = {\esss
K/\alpha} \cdot \delta(\gamma - \gamma^{\ast})$, {\esss where $K$ is
defined in the spherical approximation. $\alpha=1$ in the
spherical case, and $\alpha < 1$ for cylindrical or pancake-like
volumes reproducing the current sheet geometry.}
{\corr The peak photon flux above 100 MeV is measured in units of
$ 10^{-8} \rm \, ph \, cm^{-2} \, s^{-1}$}.
{\corr These parameters are obtained,
by means of a multi-parameter fit, from the following quantities (in the observer frame):
the position of the peak photon energy, $E_p \propto \delta \gamma^{\ast 2} B$,
the peak emitted power $\nu F \propto \delta^4 K/\alpha \, l^3 B^2 \gamma^{\ast 2}$,
the rise time $\tau_1 = l/(c \delta)$, and the cooling time
$\tau_2 = 8.9 \times 10^3/[(B/\rm Gauss)^2 \, \gamma^{\ast 2} \delta]$, assuming $\delta=1$ (see text).}}
\end{table*}

\begin{figure*}
\begin{center}
\vspace*{-0.3cm} \hspace*{-0.7cm}
 \includegraphics[width=16cm]{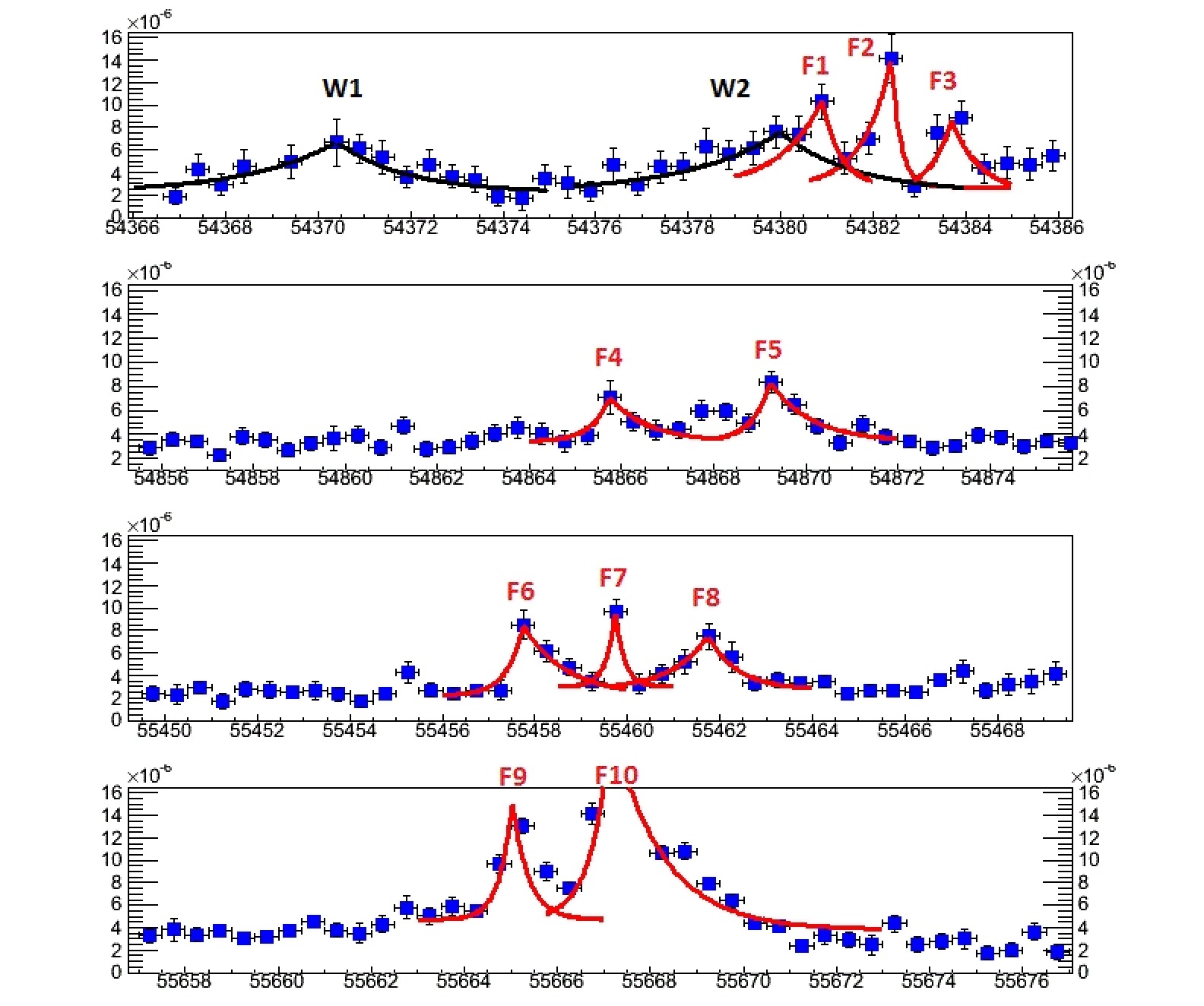}
\caption{Gamma-ray lightcurves above $100$ MeV (12-hr time bins)
from the Crab (pulsar plus Nebula) detected by AGILE and
Fermi-LAT. From top to bottom, the Sept. - Oct. 2007 event (AGILE
data), the Feb. 2009 event (Fermi-LAT data), the Sept. 2010 event
(Fermi-LAT data
), and the Apr. 2011
event (Fermi-LAT data
).}
\label{all_events}
\end{center}
\end{figure*}

\begin{figure*}
\begin{center}
\vspace*{-0.3cm} \hspace*{-0.7cm}
\includegraphics[width=17cm]{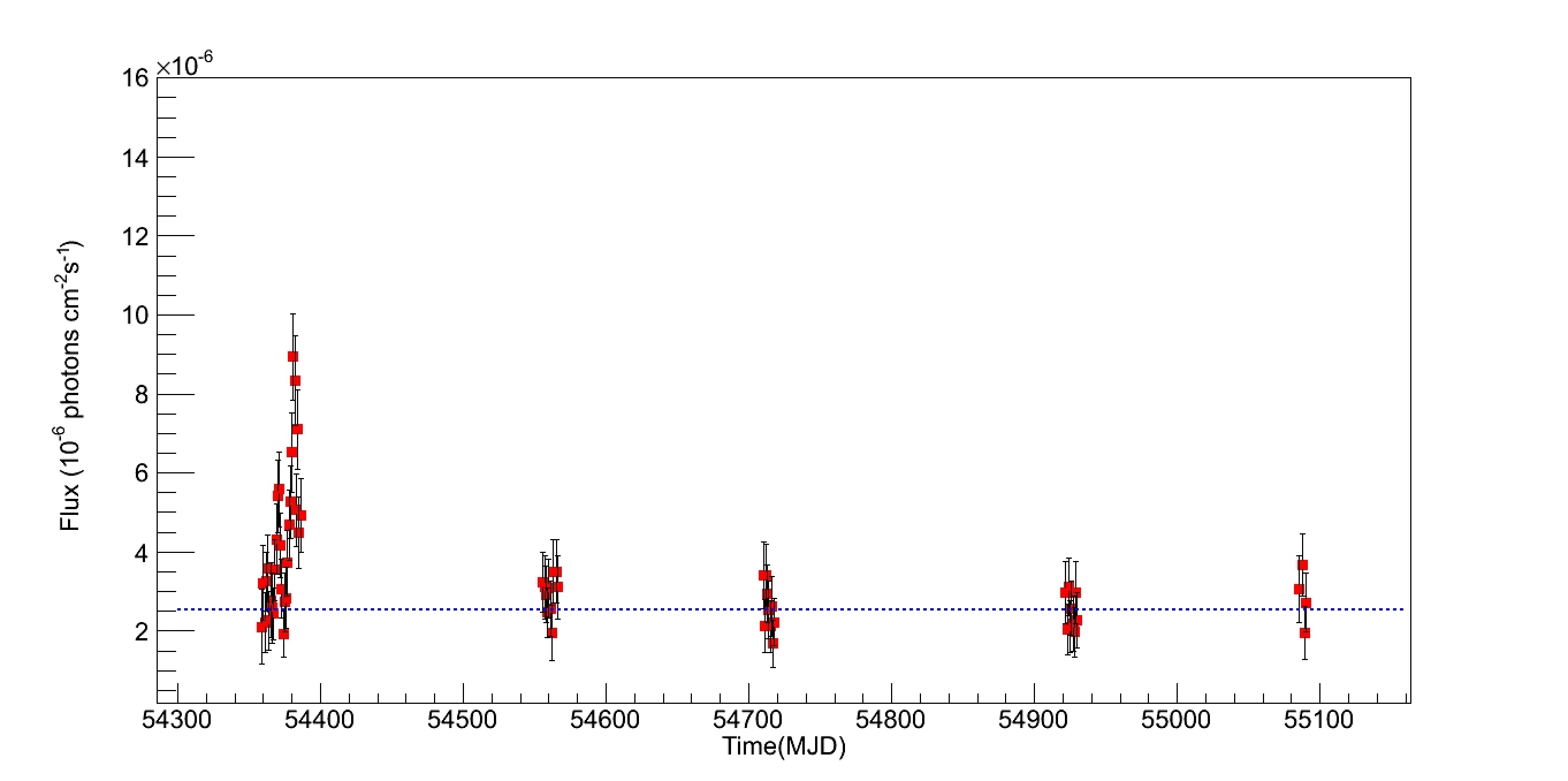}
\caption{Gamma-ray {\eess 1-day binned} lightcurve above $100$ MeV of the five AGILE
observations of the Crab Nebula {\mtt in} pointing mode, from
Sept. 2007 to Oct. 2009. {\mtt The gamma-ray flux enhancement
during the September-October 2007 pointing (MJD = 54366 \-- 54386) is
evident.}} \label{AGILE_LC}
\end{center}
\end{figure*}

\begin{figure*}
\begin{center}
\vspace*{-0.3cm} \hspace*{-0.7cm}
\includegraphics[width=17cm]{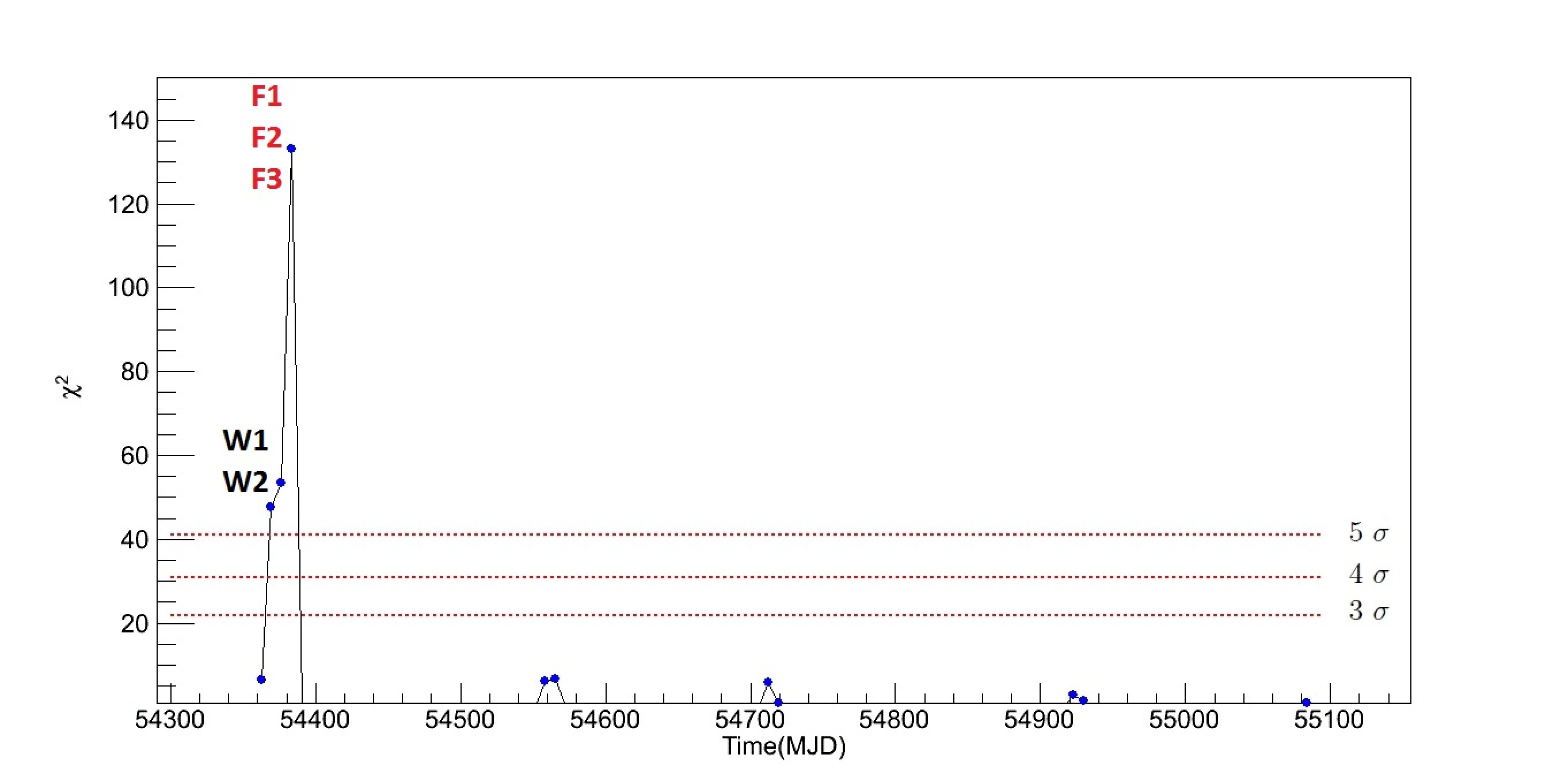}
\caption{{\eess Plot of the $\chi^2$ values
(each calculated for 7-day time intervals based on 1-day binned data) as a function
of time for the AGILE data on the Crab covering the period Sept. 2007/Oct. 2009 in pointing mode.} The red dashed lines indicate
the $\chi^2$ corresponding to the 3$\sigma$, 4$\sigma$ and
5$\sigma$ confidence level.} \label{AGILE_CHI2}
 \end{center}
 \end{figure*}

 \begin{figure*}[t!]
\begin{center}
\vspace*{-0.3cm} \hspace*{-0.7cm}
 \includegraphics[width=16cm]{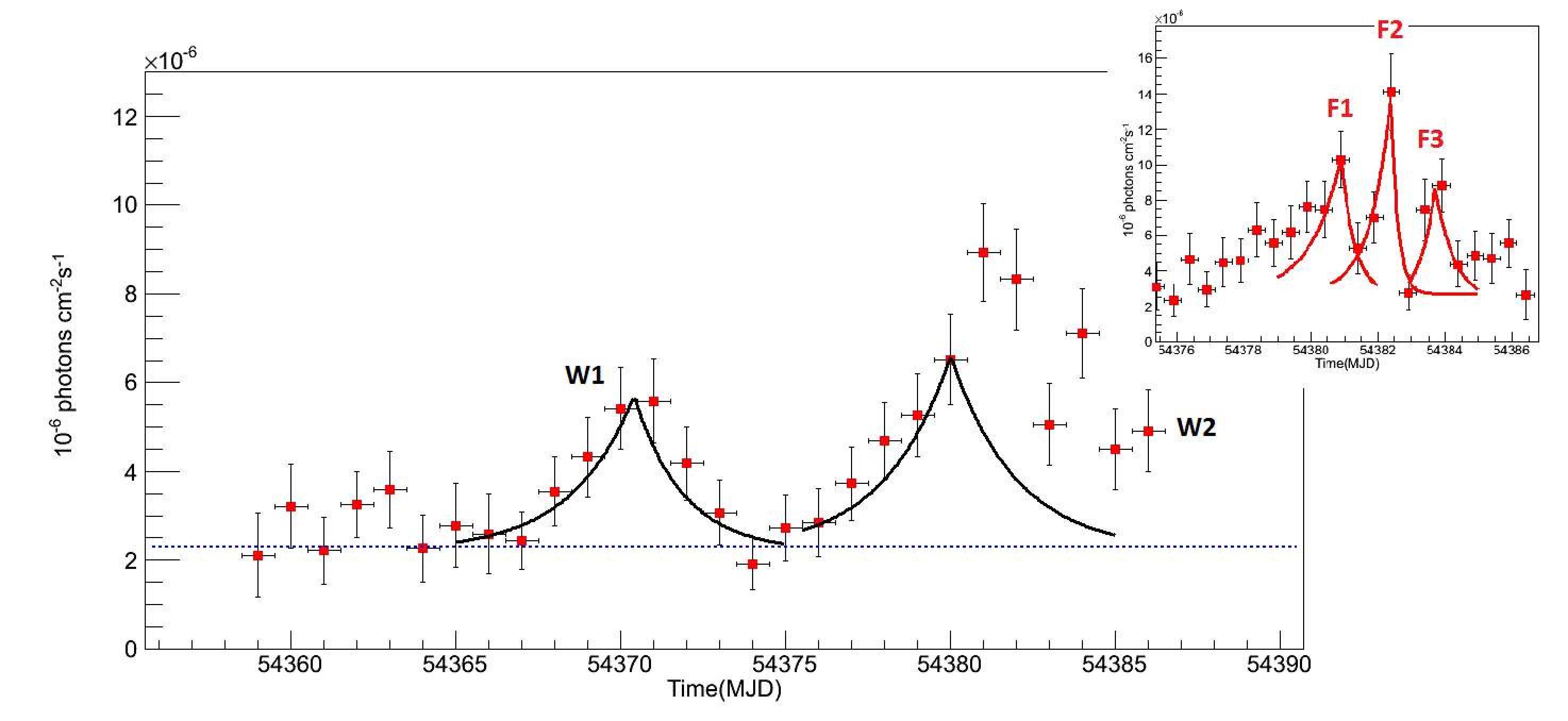}
\caption{Lightcurve (1-day bin) of the Sept.- Oct. 2007 Crab Nebula
flare detected by AGILE. {\es In the inset the 12-hr bin
lightcurve around the flare.} This episode is characterized by a
very strong variability, with waves (black line, marked with a
\emph{W}) and flares (red line, marked with an \emph{F}.)}
\label{flare_2007}
 \end{center}
 \end{figure*}

\newpage

\section{Overview of the main gamma-ray flares}

{\mtt Four major episodes of intense gamma-ray flaring from the
Crab Nebula have been detected by AGILE and Fermi-LAT
\cite{tavani3,abdo2,striani11,vittorini,2012ApJ...749...26B}.
The definition of a ``flare'' adopted in this paper is that of a
single gamma-ray enhancement event with a risetime $\tau_1 < 1$
day and flux $ F > 700 \times 10^{-8} \rm \, ph \, cm^{-2} \, s^{-1}$ above 100 MeV.
Table 1 summarizes the flaring  events that we find by considering
the AGILE and the publicly available Fermi-LAT database.
{\mtta These events show a complex time structure, being composed
of several sub-flares} that we classify as $Fn'$, with $n'$ a
progressive number. Fig. 1 summarizes the four major flaring
episodes  with the same temporal and flux scales. The colored
curves are indicative of the flaring behavior that in most cases
can be represented by {\eess an exponential fit\footnote{We use the fitting function
$$ f=\left \{\begin{array}{ll} {\corr F_b} + A
\exp(-|t-t_{P}|/\sigma_{1}) \hbox{  for  } t\leq t_{p}
\\{\corr F_b} + A \exp(-|t-t_{p}|/\sigma_{2}) \hbox{  for  } t> t_{p}
\end{array}
 \right .
$$ where {\corr $F_b$}
is the flux baseline, $A$ a flux parameter,
$t_p$ is the peak time, $\sigma_{1}$ and $\sigma_{2}$ the rise and
decay time constants. The rise and decay times, half to maximum
amplitude, are obtained as $\tau_{1,2} = [\ln(2)] \, \sigma_{1,2}
$.}}(e.g., Norris et al. 1996), characterized by rising ($\tau_1$)
and decay ($ \tau_2$) {\mtta timescales} as given in Table 1. We
also report in Table 1 the data fitting physical parameters
(average local magnetic field $B$, typical particle Lorentz
factor $\gamma^{\ast}$, {\corr and characteristic size $l$ of the emitting region}
of the flaring episodes). {\corr We determine these parameters
and their uncertainties from the time constants $\tau_1$ and $\tau_2$} (see discussion below).
Whenever applicable (2010 and 2011 events),
the AGILE and Fermi-LAT data are consistent both in flux and spectral properties. }

Motivated by the claim of fast variability in the September 2010
event (Balbo et al. 2011) and the dramatic detection of the short
timescale variability in the Crab gamma-ray flare of April 2011
(Buehler et al. 2011), we revisited the analysis of the Crab flare
detected by Fermi in Feb. 2009, and the AGILE and Fermi analysis
of the September 2010 event (12-hr bin lightcurve for $E > 100$
MeV, first two panels of Fig.~\ref{all_events}). This revised
analysis shows for both the 2009 and  2010 events a sequence of
three flares, that could not be previously appreciated with a
2-day time bin. Regarding the September 2010 event, both the AGILE
data and the Fermi-LAT data agree quite well, and are {\mtta
consistent} with the analysis presented in Balbo et al. 2011.

\section{The September-October 2007 event detected by AGILE}

The Crab pulsar plus Nebula is a primary source for gamma-ray calibration, and AGILE pointed
at the source several times
{\corr during the pointing mode phase from July 2007 until Oct. 2009
for different geometries and off-axis
angles.} The ideal  periods {\mtta during the year for AGILE
pointings of the Crab region are}  September-October and
March-April as determined by the solar panel constraints. Fig.
\ref{AGILE_LC} shows the overall gamma-ray lightcurve collecting
all the available AGILE observations which includes {\mtta
several} pointings in 2007, 2008 and 2009. Typically, pointing
observations lasted $\sim$10 days, except for the initial
pointing in September-October 2007 that was substantially longer.

During this first pointing, indeed, we detected an episode of
enhanced gamma-ray emission. A preliminary lightcurve of this
episode, performed with a 1-day bin, was presented in Tavani et
al. 2011a. In that study we reported a quite long episode (about 2
weeks) of enhanced emission with a 1-day  peak flux of
$F_p=(890\pm110)\times 10^{-8} \rm \, ph \, cm^{-2} \, s^{-1}$
above 100 MeV.

Fig.~\ref{AGILE_LC} shows the 1-day lightcurve ($E > 100 $MeV) of
the Crab (pulsar plus Nebula) extended to the five observations of
the Crab Nebula performed by AGILE during the pointing mode. {\es
In order to search for {\mtt  statistically significant enhanced
gamma-ray emission on  timescales of  weeks,} we calculated from
the data of Fig.~\ref{AGILE_LC} the 7-day $\chi^2$ curve\footnote{ For the AGILE data, we started with the 1-day binned
flux data, and then calculated the $\chi^2$ values summing over
7 days, $\chi^2=\sum_{i=1}^7 \frac{(F_{o}(i) -
F_{s})^2}{\sigma_i^2}$, where $F_{o}(i)$ is the i-th observed
flux, $F_{s}$ is the steady state flux, and $\sigma_i$ is the i-th
flux error.}} (Fig.~\ref{AGILE_CHI2}) in the null hypothesis that
the Crab Nebula is constant at its average flux ($ F_s = 220
\times 10^{-8} \rm \, ph \, cm^{-2} \, s^{-1}$ in the
Catalog of Pittori et al. 2009).

We then calculated the probability of obtaining a given $\chi^2$
(considering the degrees of freedom) in the null hypothesis. If
$\chi^2_{obs}$ is the observed value of $\chi^2$, {\corr p($\chi^2$,n)} is
the probability of obtaining a value of $\chi^2 \geq \chi_{obs}$,
with $n$ degrees of freedom. {\mtt For 7 degrees of freedom, } the
{\mtt horizontal} dashed lines in Fig.~\ref{AGILE_CHI2} indicate
the value of $\chi^2$ (quantiles) corresponding to the $3\sigma$,
$4\sigma$ and $5\sigma$ confidence level, {\mtt respectively.} We
see that the {\mtt post-October, 2007 } four observations of the
Crab Nebula performed by AGILE in the pointing mode are compatible
with the average emission within $1\sigma$. {\mtt However,  }
three {\mtt  episodes during} the first observation (from MJD
$\simeq$ 54360 to MJD $\simeq$ 54395) are above $5 \sigma$ with
respect to the Crab Nebula average emission.

{\mtta The lightcurve extending for} 20 days including the first
pointing, from Sept. 24, 2007 to Oct. 13, 2007, {\mtta is}
shown in Figure~\ref{flare_2007}. {\es In the inset we show a zoom
of the lightcurve focused on the short variability episodes.}
{\mtta Three events stand out above 5 sigma (for a 7-day time bin)
in Fig. 3. The most significant corresponds to a flaring sequence,
and two other events can be attributed to a less intense but long
timescale enhanced emission anticipating the flares,} with an
average flux of $\sim {\eees450} \times 10^{-8} \rm \, ph \, cm^{-2} \,
s^{-1}$ and a rise and decay time of the order of {\mtt several}
days.

{\mtta These slow  components of enhanced \ggg emission show
features different than those of flares that typically have} rise
and decay times of the order of 12-24 hr, and peak fluxes ranging
from $F_{p,5} \simeq 800  \times 10^{-8} \rm \, ph \, cm^{-2} \, s^{-1}$
up to $F_{p,10}\simeq 2500  \times 10^{-8} \rm \, ph \, cm^{-2} \, s^{-1}$
{\mtt (as for {the Crab super-flare of Apr. 2011)}}.

{\mtt These episodes of slow enhanced emission are significant
(see discussion below):
 we call them \emph{waves}.
We indicate with $W_1$ the emission from MJD $\simeq$ 54367 to
MJD $\simeq$ 54374. From MJD $\simeq$ 54376 to MJD $\simeq$ 54382 we
observe another {\mtta event} of enhanced emission, that we
interpret as a second ``wave'' with a small flare superimposed at
MJD $\simeq$ 54381. We name this second region  $W_2$. {\es The
12-hr lightcurve in the inset of Fig.~\ref{flare_2007} shows that
the 2007 {\mtt peak intensity
event, that in our previous 
analysis \cite{tavani3} appeared unresolved}, 
is actually composed of three different {\mtt flaring components},
that we indicate by $F_1$, $F_2$, and $F_3$}.
{\corr Peak fluxes, rise times and decay times (estimated with an exponential fit)
 for $F_1$, $F_2$, and $F_3$ and for $W_1$ and $W_2$ are presented in Tab.~\ref{tab1} and Tab.~\ref{tab2}.}
{\corr In order to take into account the number of 1-day maps (trials)
carried out in our search for enhanced emission,}
{\es we also calculated the post-trial significance for $W_1$ and
$W_2$, where the post-trial probability is given by
$P_{post}=1-(1-p)^{N_t}$ {\mtt with} $p$ the pre-trial
probability {\corr obtained from the $\chi^2$ test},
and $N_t$ the number of trials. AGILE {\mtt in
pointing mode} observed the Crab for a total time of $\sim 50$
days; considering that the $\chi^2$ {\mtt value} is calculated
over 7 days, the number of trials is $N_t \simeq 7$. {\mtt  We
find that the post-trial significance is $P_{post}> 5\sigma$ for
both $W_1$ and $W_2$.}

\begin{figure}[t!]
\begin{center}
 \includegraphics[width=9.0cm]{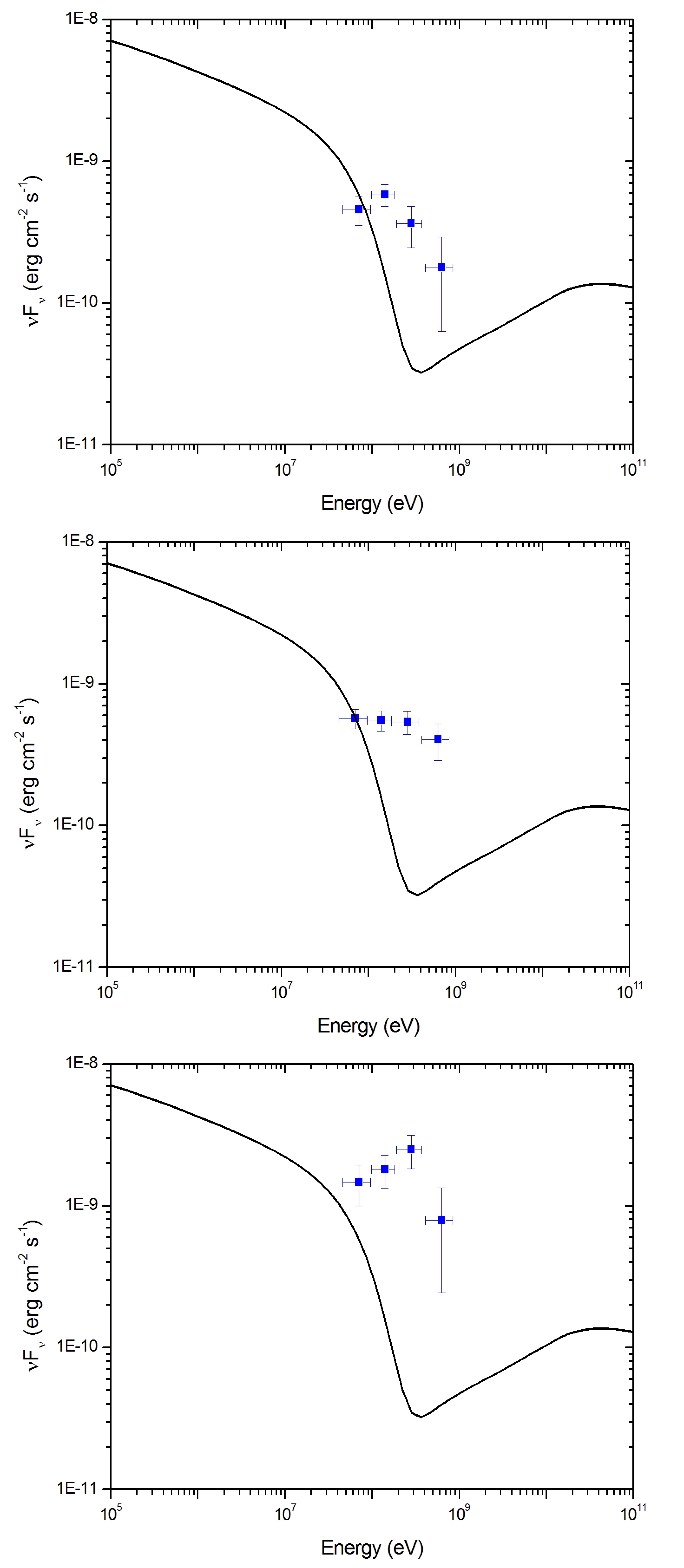}
\caption{AGILE-GRID gamma-ray pulsar-subtracted spectrum of the
Crab Nebula $W_1$ (integrating from MJD = 54368 to MJD = 54374, top
panel), $W_2$ (integrating from MJD = 54376 to MJD = 54381, central
panel) and the major flare $F_2$ (12-hr integration, from
MJD = 54382.5 to MJD = 54383.0, bottom panel).} \label{spectra}
 \end{center}
 \end{figure}


{\mtt The spectral properties of the $W_1$ and $W_2$ waves and of
the flares in September-October 2007 are interesting.}
Fig.~\ref{spectra} shows the spectral analysis of $W_{1}$
(obtained integrating from MJD = 54368 to MJD = 54374), $W_{2}$
(obtained integrating from MJD = 54376 to MJD = 54381), and the
spectral analysis of $F_2$ obtained with a 12-hr integration. (We
notice that the spectrum of $W_{2}$ might be contaminated by
presence of the first flare, $F_1$).


{\corr We find that the differential} particle energy distribution
function (per unit volume) can be described by a monoenergetic function,
 $d \, n / d \, \gamma
= {\esss K/\alpha} \cdot \delta(\gamma-\gamma^{\ast})$, where $\gamma$
is the particle Lorentz factor, {\eees and} $\gamma^{\ast}$ is the monochromatic
value of the particle energy. {\mtta The constant $K$} is defined
in the spherical approximation,
 with $\alpha=1$ in the spherical case,
  and $\alpha < 1$ for cylindrical or pancake-like volumes.
 A power-law distribution and/or a relativistic
Maxwellian distribution were {\corr also} shown to be consistent with the
flaring data (Tavani et al., 2011a, Striani et al. 2011b). We adopt
{\mtta here} a monochromatic distribution that deconvolved with
the synchrotron emissivity leads to an emitted spectrum
practically indistinguishable {from} the relativistic Maxwellian
shape (e.g., Striani et al. 2011b, Buehler et al. 2012). This
distribution is in agreement with all available gamma-ray data
(for both flaring and ``wave'' behavior, see below), and reflects
an important property of the flaring Crab acceleration process
(Tavani 2013). }
{\eess In our model we have five free {\corr physical} parameters:
{the Lorentz factor} $\gamma^{\ast}$, the local magnetic field $B$, the electron density {\corr constant } $K$,
the dimension of the emitting region $l$, and the Doppler factor $\delta$.
The values of these parameters are obtained,
{\eees by means} of a multi-parameter fit, from the following quantities (in the observer frame):
the position of the peak {\corr photon energy}, $E_p \propto \delta \gamma^{\ast 2} B$,
the peak {\corr emitted power} $\nu F \propto \delta^4 K/\alpha \, l^3 B^2 \gamma^{\ast 2}$,
the rise time $\tau_1 = l/(c \delta)$, and {\corr the cooling time
$\tau_2 = 8.9 \times 10^3/[(B/\rm Gauss)^2 \, \gamma^{\ast} \delta]$.}}
We fix the Doppler factor at the value $\delta=1$ {\mtt (relaxing this
condition leads to slightly different constraints for $B$ and
$\gamma^{\ast}$ that can be easily calculated without altering the main
conclusions of our paper).} {\corr We determine the characteristic timescales
$\tau_1$ and $\tau_2$ by a 3-parameter model (see note 1),
and from those values we deduce the other physical quantities.}
{\corr We find for $W_1$ a magnetic field
$B=(0.8\pm0.2)$ mG, a Lorentz factor $\gamma^{\ast}=(4\pm1) \times 10^9$,
and a typical emitting {\esss
length range $l = (0.5 \-- 1.5) \times 10^{16}$ cm}. For $F_2$,
we find $B = (1.5\pm0.5$) mG, $\gamma^{\ast} = (3\pm1)\times 10^9$,
and $l = (1.5\pm 0.7) \times
10^{15}$ cm.}
{\mtt In our model, the total number of accelerated particles
producing the gamma-ray wave/flaring behavior is in the range
{\esss $N \sim (1 - 3)\cdot 10^{38} (\Delta \Omega/4\pi$)},
{\mtta with $\Delta \Omega$ the solid angle of the \ggg emission}.
It is interesting to note that besides the differing values of the
magnetic field and particle densities, the typical Lorentz factor
{\corr and total number of radiating particles are} similar for the ``wave'' $W_1$ and the flare $F_2$. We
find that this {\eess is} a typical behavior of the transient gamma-ray
emission that appears to be well represented by a {\corr monochromatic
particle distribution function with $\gamma^{\ast} \simeq (3\--5) \times
10^9$. }}





\section{Search for enhanced gamma-ray emission in the Fermi-LAT data}

Motivated by the waves found in the AGILE data, we searched for a
similar type of enhanced gamma-ray emission in the {\mtt publicly
available} Fermi-LAT data. Fig.~\ref{Fermi_LC} shows the
lightcurve of the Crab (pulsar + Nebula) during the period Sept.
2008 - May 2012 obtained {\mtt by} a standard unbinned likelihood
analysis of the Fermi data ({\es 2-day bin}). The three major
flares from the Crab Nebula detected by Fermi-LAT, and the {\mtta
new low-intensity event} recently announced in {\es Jul. 2012 } in
ATel $\# 4239$ {\mtta (Ojha et al. 2012)} are recognizable at
MJD $\sim$ 54869, MJD $\sim$ 55459, MJD $\sim$ 55667 and MJD $\sim$ 56113.

\begin{figure*}
\begin{center}
\vspace*{-0.3cm} \hspace*{-0.7cm}
 \includegraphics[width=17cm]{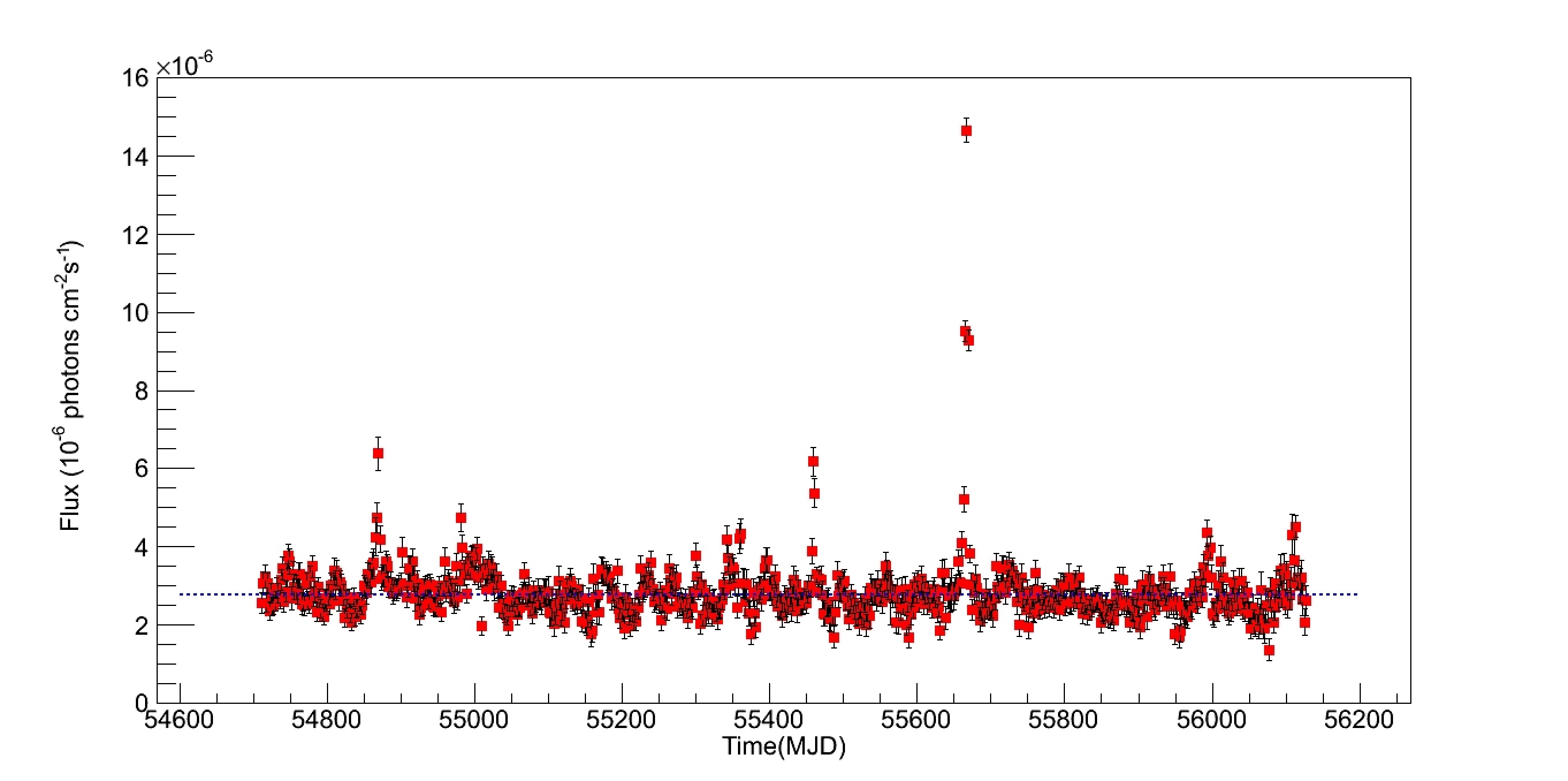}
\caption{\mtt Fermi-LAT 2-day bin lightcurve above 100 MeV of the
Crab (pulsar plus Nebula) spanning the time period
Sept. 2008/Jul. 2012. Results obtained with the publicly available
unbinned likelihood analysis software.} \label{Fermi_LC}
 \end{center}
 \end{figure*}
\begin{figure*}
\begin{center}
\vspace*{-0.3cm} \hspace*{-0.7cm}
 \includegraphics[width=17cm]{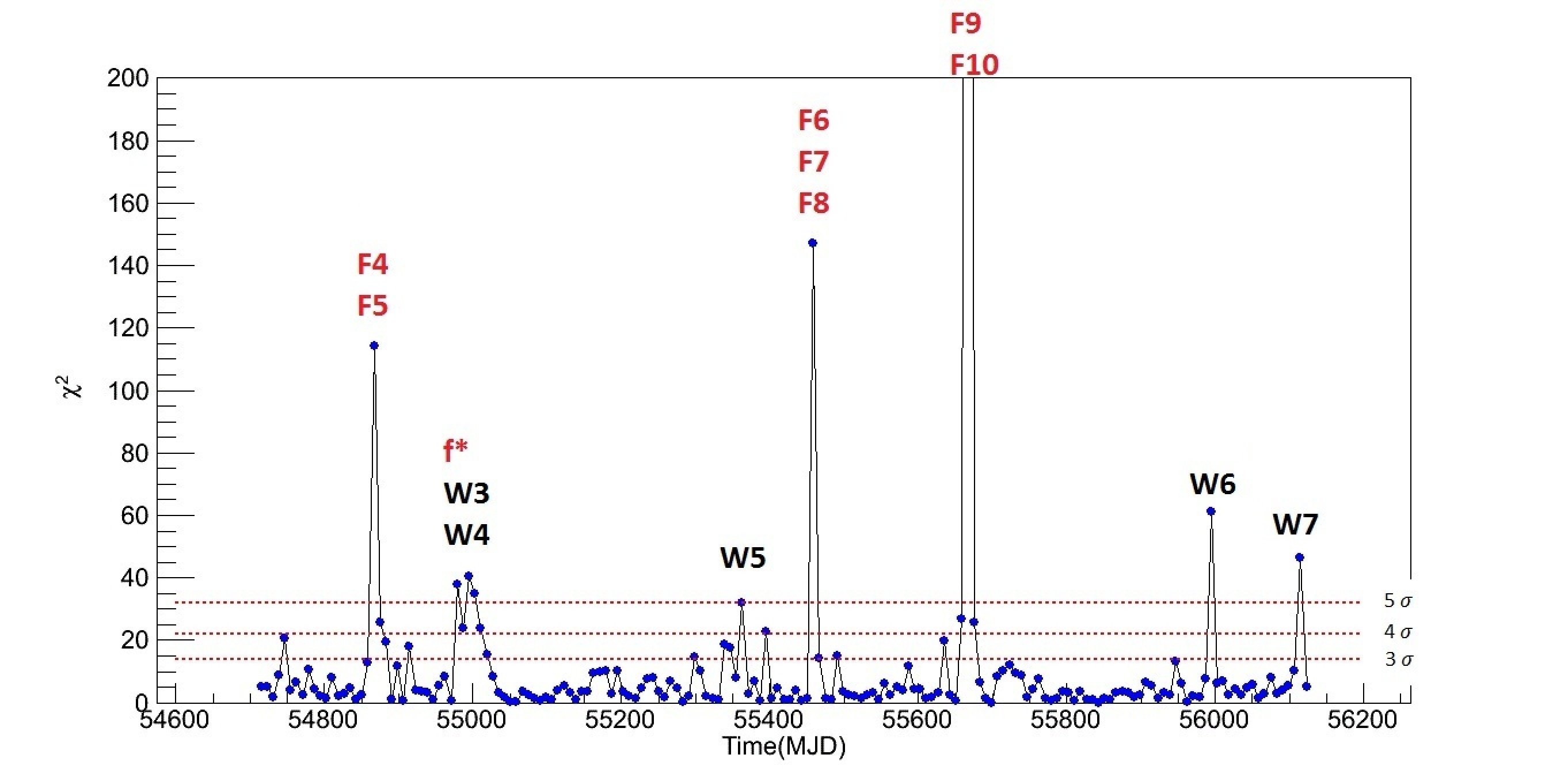}
\caption{{\mtt Plot of the $\chi^2$  {\eess values (each calculated for 8-day time intervals,
based on 2-day binned data) as a function of time for the gamma-ray Fermi-LAT data
on the Crab covering the period Sept. 2008/Jul. 2012}}.}
\label{Fermi_CHI2}
 \end{center}
 \end{figure*}
\begin{figure*}
\begin{center}
\vspace*{-0.3cm} \hspace*{-0.7cm}
 \includegraphics[width=19.5 cm]{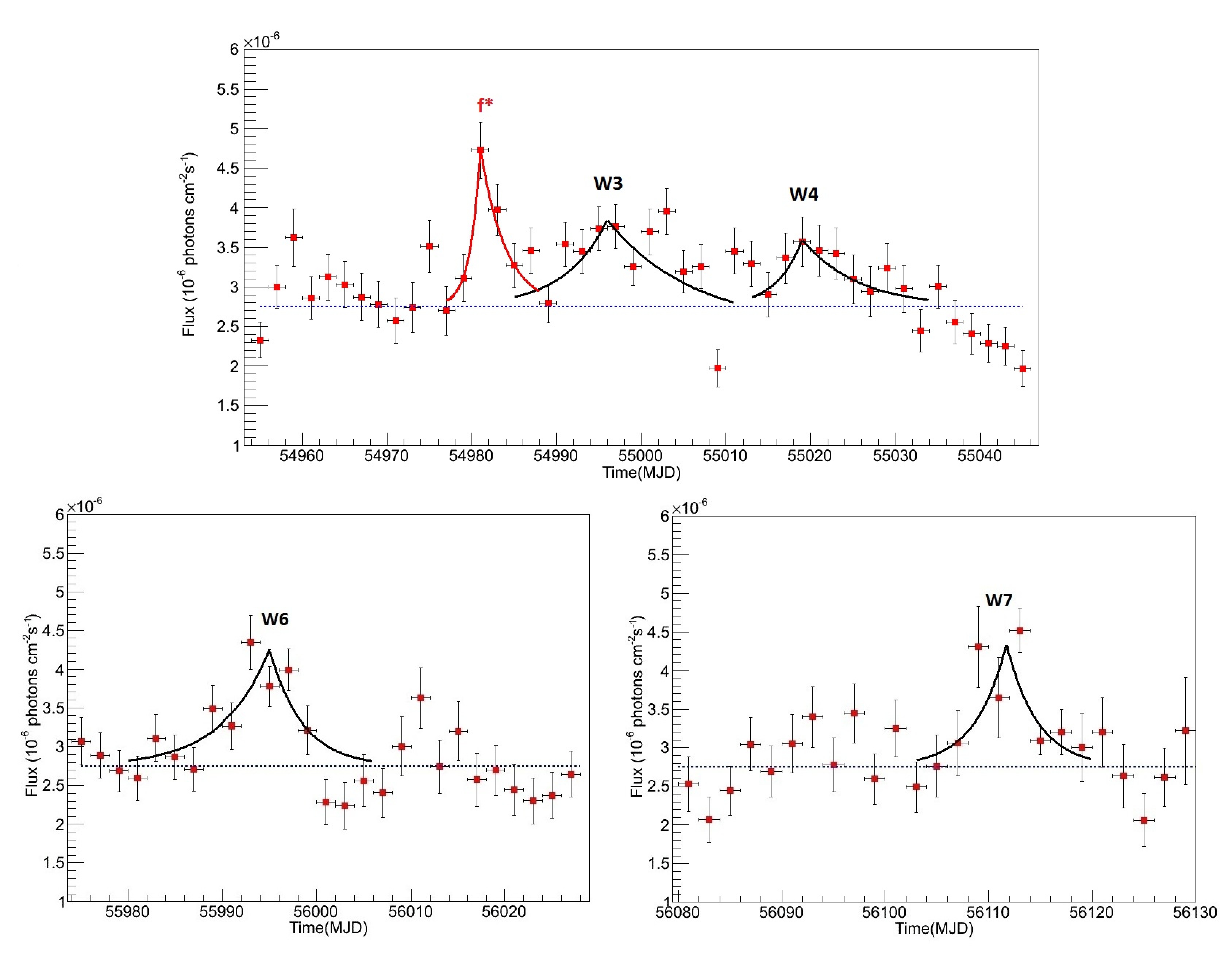}
\caption{ \mtt The most prominent \emph{wave} episodes in the
Fermi-LAT data with more than $5\sigma$ enhancements  above the
Crab average emission. {\eess 2-day binned gamma-ray flux values
(in unit of $10^{-6} \, \rm ph \, cm^{-2} \, s^{-1}$) above 100
MeV as a function of time. We notice that the event marked as $f^{\ast}$ is intermediate between flares and waves.}} \label{Fermi_5sigma}
 \end{center}
 \end{figure*}

{\mtta Starting with the 2-day bin  
lightcurve
we  calculated the $\chi^{2}$ distribution based on  8-day
integrations, and tested the null hypothesis for a source with
 a constant average flux }  
 $ F_F =  (296 \pm 2.5) \times 10^{-8} \rm \, ph \, cm^{-2} \, s^{-1} $
above 100 MeV for the pulsar plus Nebula signal}\footnote{See,
e.g., http://www.asdc.asi.it/fermi2fgl.}. {\mtt We used the same
procedure employed for the study of the AGILE data\footnote{For the Fermi-LAT data, we started with the 2-day binned
flux data, and then calculated the $\chi^2$ values summing over
8 days, $\chi^2=\sum_{i=1}^4 \frac{(F_{o}(i) -
F_{s})^2}{\sigma_i^2}$, where $F_{o}(i)$ is the i-th observed
flux, $F_{s}$ is the steady state flux, and $\sigma_i$ is the i-th
flux error.}}. {\es In Fig.~\ref{Fermi_CHI2} we show the 8-day
bin $\chi^2$ calculated for the whole Fermi-LAT dataset ($\sim 4$
years).} {\mtt As} in Fig. ~\ref{AGILE_CHI2}, the dashed lines in
Figure ~\ref{Fermi_CHI2} indicate the $\chi^2$ corresponding to
the $3\sigma$, $4\sigma$ and $5\sigma$ confidence level {\mtt for
4 degrees of freedom}. {\mtt In addition to  the three major peaks
of the $\chi^2$ curve } (corresponding to the  major flares
of Feb. 2009, Sept. 2010 and Apr. 2011)
{\mtt we identify several episodes of enhanced gamma-ray emission
(above a $5 \sigma$ pre-trial significance) that we call $W_3, W_4, W_6$ and $W_7$
 in Fig.~\ref{Fermi_5sigma}. These episodes have apparent
durations in the range from 8 to 50 days (see the complex marked
as $W_3-W_4$), and an average flux in the range } $F =
(350-500)\times 10^{-8} \rm \, ph \, cm^{-2} \, s^{-1}$.
The {\mtta event  $W_7$ is coincident with the {\mtt very recent
enhancement episode}  detected\footnote{Due to solar panel
constraints, this event was unobservable by AGILE.} by Fermi-LAT
(Ojha et al., 2012).}
{\mtt Fig.~\ref{Fermi_5sigma} shows the detailed lightcurves of
the Fermi-LAT wave episodes with the largest post-trial
significance\footnote{{\es The post-trial significance was
calculated as in the previous {\mtta Section}. Considering four
years of Fermi data, and a {\mtta 8-day time bin}, the number of
trials {\mtta turns out to be} $N_t \simeq 180$.}}}.

{\mtt Fig.~\ref{Fermi_5sigma}  shows a remarkable episode of
{\mtta ``wave''} enhanced emission near MJD = 55000}. The Crab was
for about 50 days above $5\sigma$ from its standard gamma-ray
flux, with an average flux in this period $F = (340 \pm 6)
\times 10^{-8} \rm \, ph \, cm^{-2} \, s^{-1}$. {\mtta This event
{\mtta appears} quite complex\footnote{{\mtt For simplicity, we
use {\eess an exponential} approximation for the ``wave'' emission. Admittedly,
{\eess this} approximation for the episode of Fig.~\ref{Fermi_5sigma}
centered on MJD $= 55000$ is not adequate. It is shown in
Fig.~\ref{Fermi_5sigma} for illustrative purposes only.}}. A minor
flare (that we mark as $f^{\ast} $, at MJD $\sim$ 54982)
anticipates a complex and long emission that we approximate as two
``waves'', $W_3$ and $W_4$.}
{\mtta The ``waves'' $W_3$ and $W_4$ last} for $\sim 15-20$ days
{\mtta each}, and show a post-trial significance (on a 8-day
timescale) {\mtt near or} above $5\sigma$ {\mtta (see also Table 3
in the Appendix)}. The total {\mtt episode} that includes
$f^{\ast} $, $W_3$ and $W_4$ has a time duration of $\sim 50$
days, a pre-trial probability $p=2\times 10^{-25}$ and a
post-trial significance $\sigma_{post}>10$.
{\es For each ``wave'' that we found at $5\sigma$ above the Crab
average emission, we estimated the rise $\tau_1$ and the decay
time $\tau_2$, the average flux, the peak flux, the probability of
obtaining the given $\chi^2$ in the null hypothesis, and the post-trial
significance. The rise and decay timescales are estimated
{\eess with an exponential fit.}
The results are summarized in Tab.~\ref{tab2}.

\begin{table*} 
\begin{center}
  \small
   \caption{Table of the \emph{waves} above $5\sigma$ post-trial from the Crab average emission found in the AGILE and Fermi data.}
  \begin{tabular}{|c|c|c|c|c|c|c|c|c|}
  \hline
    Name & MJD & Duration & $\tau_1$ & $\tau_2$ & Average Flux & Peak Flux &  Pre-trial & Post-trial \\
   &  & (days) & (days) & (days) & $ (10^{-8} \rm \, ph \, cm^{-2} \, s^{-1})$  &  $ (10^{-8} \rm \, ph \, cm^{-2} \, s^{-1})$   & p-value & significance\\ \hline
  $W_{1}$ & 54368-54373     & 5      &  {\eess $2\pm 1$}   & {\eess $2\pm 1 $}     &$440\pm40$  &$670\pm200$   &$4.5 \times 10^{-8}$   & 5.0 \\\hline
  $W_{2}$ & 54376.5-54382.5 & 6      & {\eess $2.5\pm 1$}  & {\eess $2\pm 1 $}     &$480\pm40$  &$760\pm140$   &$3.0 \times 10^{-9}$   & 5.5\\ \hline
  $W_{3}$ & 54990-55008     & 18     & {\eess $5\pm 2.5$}    & {\eess $10\pm 5$}    &$352\pm9$   &$380\pm30$    &$1.0 \times 10^{-8}$    & 4.6\\ \hline
  $W_{6}$ & 55988-56000      & 12     & {\eess $5\pm 2.5$}    & {\eess $3.5\pm 1.5$}   &$367\pm12$  &$435\pm35$    &$1.8 \times 10^{-12}$    & 6.2\\ \hline
  $W_{7}$ & 56108-56114      & 6      & {\eess $3\pm 1.5$}    & {\eess $3\pm 1.5$}     &$431\pm22$  &$450\pm30$    &$1.9 \times 10^{-9}$   & 5.9\\ \hline
  \hline
\end{tabular}
\end{center}
\noindent {Photon fluxes are obtained for $E_{\gamma}> 100$ MeV.}
  \label{tab2}
\end{table*}

\section{Constraints from a  synchrotron cooling model}

{\mtta We can deduce important physical parameters of the enhanced
\ggg emission by adopting a synchrotron cooling model (see also
Tavani et al. 2011a, Vittorini et al. 2011, Striani et al. 2011b).
Tables 1, 2, and 3 summarize the
relevant information for the major ``flares'' and ``waves''.}
{\corr We constrain the physical quantities of our model as described in sect. 3.}
{\mtta We find that for flares lasting 1-2 days the typical length
is  $l \simeq (1 - 2)\times 10^{15}$ cm, the density constant in the
range $K/\alpha = (2 - 8)\times 10^{-9} \, \rm cm^{-3}$, the
typical Lorentz factor $\gamma^{\ast} =(2.5\--4.5) \cdot 10^{9}$, and the
local magnetic field affecting the cooling phase in the range $B =
(1-2) $ mG. As discussed above, the total number of radiating particles,
{\corr in case of unbeamed ($\delta=1$) isotropic emission,
is \corr $N\sim (1\--3) \cdot 10^{38}$} .}

{\mtta For the ``wave'' episodes, the typical length is  $l >
10^{16}$ cm, the density constant is in the range $K/\alpha = (2 -
8)\times 10^{-11} \, \rm cm^{-3}$, {\corr the  Lorentz factor $\gamma^{\ast}
=(3-5) \cdot 10^{9}$}, and the local magnetic field in the range
$B = (0.5-1)$ mG. The total number of particles involved is of
the same order as in the case of flares.


{\mtt We show in  Fig.~\ref{gamma_B} a schematic  representation
of the physical parameter space
{\mtta ($B$ vs. $\gamma^{\ast}$)} for  the steady emission, waves and
flares that are characterized by different and distinct regions}.
{\mtta The local magnetic field for ``waves'' and ``flares'' is
definitely amplified with respect to the standard {\corr average Crab
magnetic field ($\sim 200 \mu$G)} by a
factor 5-10, reflecting the magnetic nature of the instability
producing the enhanced \ggg emission. Remarkably, we find that,
within the monochromatic approximation of the enhanced particle
energy distribution in agreement with all spectral ``wave''/flare
data, the Lorentz factor is in the range $ 2.5 \cdot 10^9 \leq
\gamma^{\ast} \leq 5 \times 10^9$ for both  ``waves'' and ``flares''.
The indication that both ``flares'' and ``waves''  have the same
characteristic Lorentz factor $\gamma^{\ast}$  suggests a limitation
of the acceleration process, most likely induced by radiation reaction.
However, the monochromatic vs. a more extended power-law nature of the ``wave''
emission requires to be tested by additional multifrequency data.

}

\begin{figure*}[t!]
\begin{center}
\hspace*{-0.7cm}
\includegraphics[width=15.5cm]{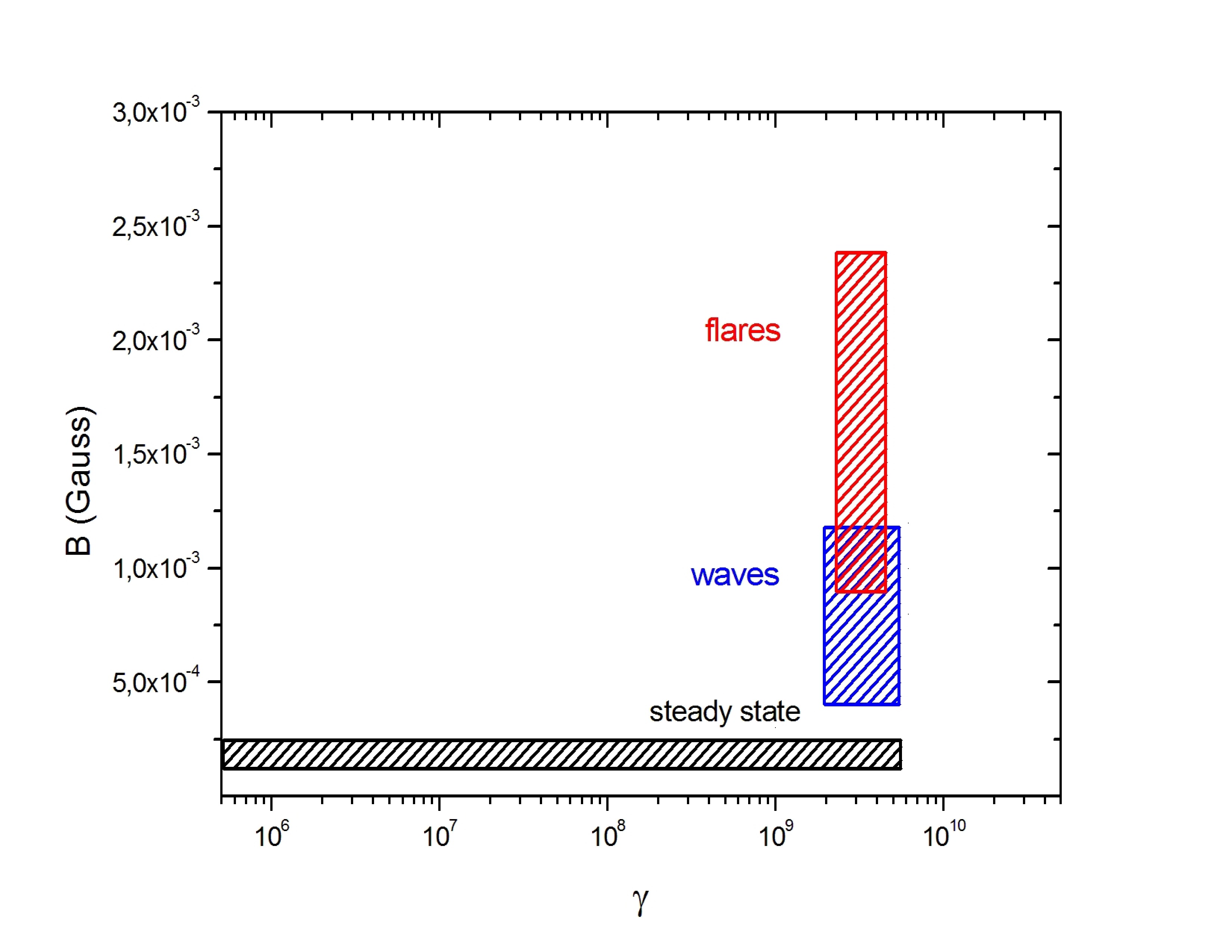}
\caption{Schematic representation of the parameter space ($B$ vs.
$\gamma$) of the Crab Nebula in different \ggg states for the
emission model adopted in this paper (monochromatic particle
energy distributions for the ``wave'' and ``flare'' states, and
power-law distribution for the steady state). For this latter
case, the shaded horizontal region indicates the range of Lorentz
factors applicable to the power-law distribution. The ``wave''
spectral shape is poorly constrained, and the effective particle
energy distribution of ``waves'' may extend to the left of the
region shaded in blue.} \label{gamma_B}
 \end{center}
 \end{figure*}

\section{Discussion and Conclusions}

In this paper we addressed the issue of the duration and intensity
of detectable enhanced gamma-ray emission from the Crab Nebula. By
considering AGILE and Fermi-LAT gamma-ray data above 100 MeV we
find that the Crab produces a broad variety of enhanced emission.
We characterize this enhanced emission as short timescale (1-2
day) ``flares'' and  long timescale (1 week or more) ``waves''.
Given the current detection level of 1-2 day enhancements (which
can be extended to longer timescales of order of 1-2 weeks), we
cannot exclude that the Crab is producing an even broader variety
(in flux and timescales) of enhanced gamma-ray emission. With the
current sensitivities of {\corr $\gamma$-ray telescopes} we can
explore an important but necessarily limited range of flux and
spectral variations.
{\corr It is interesting to note that
what we called ``flares'' and ``waves'' (a somewhat arbitrary
division) share the same spectral properties. Given the current
gamma-ray sensitivities, we could have detected different spectral
behaviors in the energy range 50 MeV - 10 GeV. Most likely, flares
and waves are the product of the same class of plasma
instabilities that we show acting on different timescales and
radiation intensities. The overall detectable transient emission
appears to be without any discernible pattern (see, e.g., Fig. 7).
Whether or not the instability driver of this process is truly
stochastic in flux and timescales will be determined by a longer
monitoring of the Crab Nebula.} The transient behavior is the
topic of current intense theoretical investigation\footnote{See
the reports presented at the meeting ``The Flaring Crab: Surprise
and Impact'', $\rm www.iasf-roma.inaf.it/Flaring\_Crab$.}
(e.g., Bednarek \& Idec 2011, Komissarov \& Lyutikov 2011, Uzdensky,
Cerutti \& Begelman 2011, Cerutti, Uzdensky \& Begelman 2012,
Bykov et al. 2012, Sturrock \& Aschwanden 2012, Clausen-Brown \&
Lyutikov 2012, Kohri et al. 2012, Lyubarsky 2012, Komissarov
2013, Mignone et al., 2013, Tavani 2013).
The pulsar wind outflow and nebular interaction conditions need to
be strongly modified by instabilities in the relativistic flow
and/or in the radiative properties. Plasma instabilities possibly
related to magnetic field reconnection in specific sites in the
Nebula can be envisioned. However, {\mtta evidence for magnetic
field reconnection events in the Crab Nebula is elusive, and } no
 optical or X-ray emission in coincidence with the gamma-ray
flaring has been {\mtta unambiguously} detected to date (e.g.,
Weisskopf et al. 2012).

Both the flaring and ``wave'' events
can be attributed to a population of accelerated electrons consistent with a
mono-chromatic or relativistic Maxwellian distribution of
typical energy $\gamma^{\ast} \sim (2.5\--5) \cdot 10^9$. The magnetic field
inducing synchrotron radiation during the decay phase of the
waves/flares is substantially larger than in the steady state, as
shown in Fig. \ref{gamma_B}. The range of particle energies
reproducing the wave/flare spectra is quite restricted with
respect to the steady state, and concentrated towards the maximum
value of the overall distribution function. This is an important
property of the flaring Crab acceleration mechanism whose maximum
energy is most likely limited by radiation reaction.

We also notice that the emitted total energies that can be deduced
for the wave ($\delta E_{\gamma,w}$) and flare ($ \delta
E_{\gamma,f}$) episodes in general satisfy the relation $\delta
E_{\gamma,w} \simeq \delta E_{\gamma,f}$. The total gamma-ray emitted energy for the wave episode
$W_1$ can be {\corr estimated as $\delta E_{\gamma,w1} \sim 10^{41} \, \rm erg$}.
During the 5 days of the $W_1$ episode, the
total Crab spindown energy {\corr is $E_{sd} \sim 2 \cdot 10^{44} \rm \,
erg$}. Therefore, the observed $W_1$ efficiency for gamma-ray
emission above 100 MeV {\corr is $\varepsilon \sim 5 \cdot 10^{-4}$}. We
expect the efficiency of synchrotron emission to be of order of
$10\%$ of the particle kinetic energy. Therefore, the
{\corr energy associated with the } wave
event $W_1$, taken here as an example of Crab ``wave'' emission,
{\corr can reach a few percent} of the total spindown energy.

We conclude that the Crab ``wave'' events are highly significant
and quite important from the energetic point of view. ``Waves''
typically imply regions larger than in the case of flares, and
smaller average magnetic fields. Their total emitted
gamma-ray energy can be {\corr comparable with} that associated with shorter
flares. More observations of this fascinating phenomenon are
necessary to improve our knowledge of the flaring Crab.

\acknowledgements

{\corr We thank an anonymous referee for his/her comments.}
Research partially supported by the ASI grants no. I/042/10/0, and
I/028/12/0.


\clearpage

\appendix

\section{Search for \ggg ``wave'' emission from the Crab Nebula}

{\mtta Table 3 shows  results of our complete analysis of ``wave''
enhanced gamma-ray emission from the Crab Nebula. Seven episodes
are identified in the AGILE/Fermi-LAT database with a pre-trial
significance larger than $5\sigma$. Table 3 provides the
corresponding p-values. When selected for post-trial significance
(see main text), only a sub-class of events survives as indicated
in Table 2.}
 {\mtta The detectability of the  ``wave'' phenomenon
produced by the Crab is clearly limited by the \ggg sensitivity
and exposure characteristics of AGILE and Fermi-LAT. Variable \ggg
emission from the Crab with timing and spectral characteristics
different from those addressed in this paper cannot be excluded.}
{\eess Fig.~\ref{Dist_chi} shows the $\chi^2$ distribution
of the Fermi-LAT gamma-ray flux data of the Crab Nebula.}

\begin{table*} [h!]
\begin{center}
  \small
  \caption{Table of the \emph{waves} above $5\sigma$ (pre-trial) from the Crab average emission found in the AGILE and Fermi data.}
    \vspace*{0.5cm}
  \begin{tabular}{|c|c|c|c|c|c|c|c|c|}
  \hline
    Name & MJD & Duration & $\tau_1$ & $\tau_2$ & Average Flux & Peak Flux &  Pre-trial & Post-trial \\
   &  & (days) & (days) & (days) & $ (10^{-8} \rm \, ph \, cm^{-2} \, s^{-1})$  &  $ (10^{-8} \rm \, ph \, cm^{-2} \, s^{-1})$   & p-value & significance\\ \hline
 $W_{1}$ & 54368-54373     & 5      &  {\eess $2\pm 1$}   & {\eess $2\pm 1 $}     &$440\pm40$  &$670\pm200$   &$4.5 \times 10^{-8}$   & 5.0 \\ \hline
  $W_{2}$ & 54376.5-54382.5 & 6      & {\eess $2\pm 1$}  & {\eess $2\pm 1 $}     &$480\pm40$  &$760\pm140$   &$3.0 \times 10^{-9}$   & 5.5\\ \hline
  $f^{\ast} $ & 54980.0-54986 & 6      &   $1\pm 0.5$      &  $2\pm1$              & $470\pm35$ & $380\pm40$  & $8.0 \times 10^{-7 }$  & 4.2 \\ \hline
  $W_{3}$ & 54990-55008     & 18     & {\eess $5\pm 2$}    & {\eess $10\pm 5$}    &$352\pm9$   &$380\pm30$    &$1.0 \times 10^{-8}$    & 4.6\\ \hline
  $W_{4}$ & 55010-55025     & 15     &  {\eess $3\pm 1$}   & {\eess $6\pm 3$}   &$326\pm10$  &$360\pm30$    &$4.6 \times 10^{-7}$   & 3.8\\ \hline
  $W_{5}$ & 55358-55362     & 4      &  {\eess $2\pm 1$}   & {\eess $2\pm 1$}   &$426\pm27$  &$430\pm30$    &$5.6 \times 10^{-7 }$  & 3.7\\ \hline
 $W_{6}$ & 55988-56000      & 12     & {\eess $5\pm 2$}    & {\eess $3\pm 1$}   &$367\pm12$  &$435\pm35$    &$1.8 \times 10^{-12}$    & 6.2\\ \hline
  $W_{7}$ & 56108-56114      & 6      & {\eess $3\pm 1$}    & {\eess $3\pm 1$}     &$431\pm22$  &$450\pm30$    &$1.9 \times 10^{-9}$   & 5.9\\ \hline
  \hline
\end{tabular}
\end{center}
  \label{tab3}
\normalsize
\noindent {Photon fluxes are obtained for $E_{\gamma}> 100$ MeV. We notice that the event marked as $f^{\ast}$ is intermediate between flares and waves.}
\end{table*}

\begin{figure}[h!]
\begin{center}
 \includegraphics[width=10.5 cm]{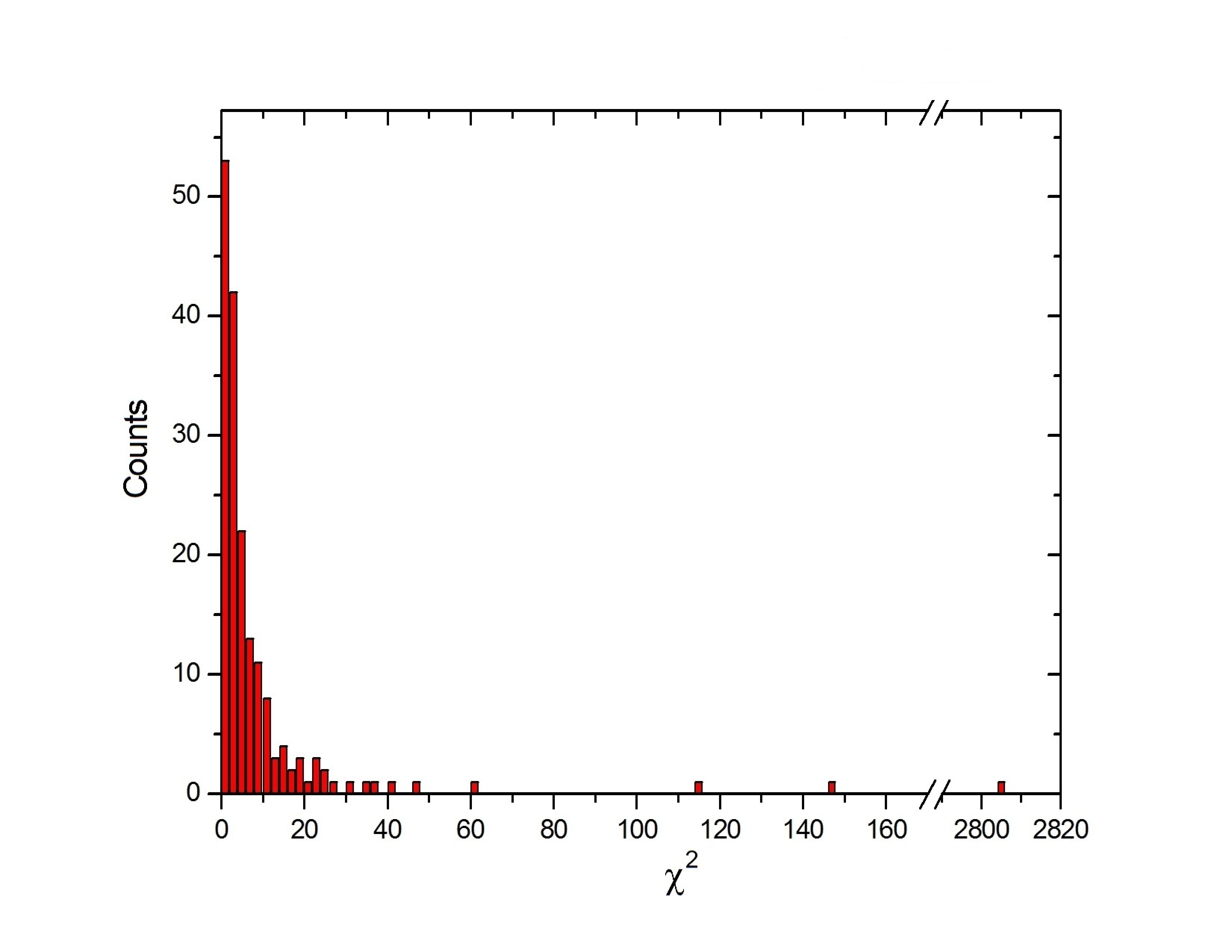}
\caption{$\chi^2$ distribution of the Fermi-LAT gamma-ray flux data
of the Crab Nebula. Each value of $\chi^2$ is integrated over 8 days.} \label{Dist_chi}
 \end{center}
 \end{figure}

\end{document}